\documentclass[12pt]{iopart}

\usepackage{graphicx}
\usepackage{hyperref}
\usepackage{verbatim}
\usepackage{xcolor} 
\usepackage{caption}
\usepackage{subcaption}
\usepackage{soul}

\definecolor{selfcommentcolor}{HTML}{FAA21A}

\definecolor{hotpink}{rgb}{1.0, 0.41, 0.71}

\newcommand{\incorporated}[2]{}

\newcommand{\PEC}[4]{{\mathop{\mathrm{PEC}}_{#1,#2\rightarrow #3}}^{\mathrm{#4}}}
\newcommand{\mean}[2]{(1/N)\sum_{#2}#1}
\newcommand{\meansquareerror}[3]{\ensuremath{\mean{\left({#1}_{#3}-{#2}_{#3}\right)^2}{#3}}}
\newcommand{\weightedmse}[4]{\ensuremath{\mean{\left(({#1}_{#3}-{#2}_{#3})/#4\right)^2}{#3}}}
\newcommand{\mserepr}[2]{\ensuremath{\mathrm{MSE}(#1, #2)}}
\newcommand{\weightedmserepr}[3]{\ensuremath{\mathrm{WMSE}(#1, #2, #3)}}

\newcommand{\neutrals}{\ensuremath{n_{\mathrm{neutral}}}}
\newcommand{\electrons}{\ensuremath{n_{\mathrm{electrons}}}}
\newcommand{\etemperature}{\ensuremath{T_{\mathrm{electrons}}}}
\newcommand{\totaldensity}{\ensuremath{n_{total}}}
\newcommand{\bulkdensity}{\ensuremath{n_{bulk}}}
\newcommand{\residualdensity}{\ensuremath{n_{residual}}}
\newcommand{\ionisationfraction}{\ensuremath{\chi}}
\newcommand{\emission}{\ensuremath{\epsilon_{\mathrm{D\alpha}}}}

\newcommand{\opticaldensity}{\ensuremath{\mu_t}}
\newcommand{\nulldensity}{\ensuremath{\mu_n}}

\newcommand{\majorantdensity}{\ensuremath{\bar{\opticaldensity}}}

\newcommand{\densityunits}{\ensuremath{\mathrm{m^{-3}}}}
\newcommand{\temperatureunits}{\ensuremath{\mathrm{eV}}}
\newcommand{\ndunits}{\ensuremath{\mathrm{n.d.}}}

\newcommand{\expten}[1]{\ensuremath{\mathrm{{10}^{#1}}}}
\newcommand{\exptwo}[1]{\ensuremath{\mathrm{{2}^{#1}}}}

\newcommand{\scinot}[1]{\ensuremath{\cdot\expten{#1}}}
\newcommand{\round}[1]{\ensuremath{\lfloor#1\rceil}}

\newcommand{\penaltyfunctional}[1]{\ensuremath{\mathrm{P}\left[#1\right]}}

\newcommand{\totalobjectivelabel}{\mathrm{total}}
\newcommand{\renderingobjectivelabel}{\mathrm{rendering}}
\newcommand{\midplaneobjectivelabel}{\mathrm{midplane}}
\newcommand{\tvaobjectivelabel}{\mathrm{TVA}}
\newcommand{\csobjectivelabel}{\mathrm{cross-section}}
\newcommand{\penaltyobjectivelabel}{\mathrm{penalty}}

\newcommand{\objectivefun}[1]{\ensuremath{L_{#1}}}

\newcommand{\totalobjective}{\objectivefun{\totalobjectivelabel}}
\newcommand{\renderingobjective}{\objectivefun{\renderingobjectivelabel}}
\newcommand{\midplaneobjective}{\objectivefun{\midplaneobjectivelabel}}
\newcommand{\tvaobjective}{\objectivefun{\tvaobjectivelabel}}
\newcommand{\csobjective}{\objectivefun{\csobjectivelabel}}
\newcommand{\penaltyobjective}{\objectivefun{\penaltyobjectivelabel}}

\newcommand{\objectiveweight}[1]{\ensuremath{w_{#1}}}

\newcommand{\renderingobjweight}{\objectiveweight{\renderingobjectivelabel}}
\newcommand{\midplaneobjweight}{\objectiveweight{\midplaneobjectivelabel}}
\newcommand{\tvaobjweight}{\objectiveweight{\tvaobjectivelabel}}
\newcommand{\csobjweight}{\objectiveweight{\csobjectivelabel}}
\newcommand{\penaltyobjweight}{\objectiveweight{\penaltyobjectivelabel}}

\begin{document}

\title[Inverse Rendering of Fusion Plasmas]{Inverse Rendering of Fusion Plasmas: Inferring Plasma Composition from Imaging Systems}

\author{
E \"Ozt\"urk$^1$\footnote{\url{https://www.imperial.ac.uk/people/ekin.ozturk17}}, 
R Akers$^2$, 
S Pamela$^2$\footnote{\url{https://stanstash.com/spamela_profile/}}, 
The MAST Team$^2$, 
P Peers$^3$\footnote{\url{https://www.cs.wm.edu/~ppeers/}} and
A Ghosh$^1$\footnote{\url{https://www.imperial.ac.uk/people/abhijeet.ghosh}}}

\address{$^1$ Department of Computing, Imperial College, Prince Consort Road, London SW7~2BZ, UK}
\address{$^2$ Culham Centre for Fusion Energy, Culham Science Centre, Abingdon OX14 3EB}
\address{$^3$ Computer Science Department, College of William \& Mary, Williamsburg, VA, 23187, USA}

\begin{abstract}
In this work, we develop a differentiable rendering pipeline for visualising plasma emission within tokamaks, and estimating the gradients of the emission and estimating other physical quantities. Unlike prior work, we are able to leverage arbitrary representations of plasma quantities and easily incorporate them into a non-linear optimisation framework. The efficiency of our method enables not only estimation of a physically plausible image of plasma, but also recovery of the neutral Deuterium distribution from imaging and midplane measurements alone. We demonstrate our method with three different levels of complexity showing first that a poloidal neutrals density distribution can be recovered from imaging alone, second that the distributions of neutral Deuterium, electron density and electron temperature can be recovered jointly, and finally, that this can be done in the presence of realistic imaging systems that incorporate sensor cropping and quantisation.
\end{abstract}
\submitto{\NF}

\maketitle

\section{\label{sec:introduction}Introduction}

In fusion research, high-speed imaging of tokamak plasma can provide useful insight about the dynamics and properties of the experiment. Accurate distributions of the neutrals, the electrons, the ions and the electron temperature allows visualisations of the plasma that matches camera images through the forward rendering of the plasma emission. Furthermore, inverting the rendering enables recovering information about the distributions from images themselves. The process of recovering a 3d representation underlying the image seen by a camera is an active research topic called inverse rendering~\cite{Tong2021} in the field of computer graphics. While methods exist for reconstructing representations such as point clouds~\cite{Niu2024}, implicit surfaces~\cite{Vicini2022sdf}, and neural representations~\cite{Kantarci2022}; for plasmas the ideal representations would be the physical quantities themselves, neutrals density, electron density and temperature, which motivates the use of inverse rendering.

Inverse rendering is a highly non-linear problem. To exploit powerful non-linear optimization algorithms, the derivatives of the objective function are needed. An active area of research in computer graphics is to leverage automatic differentiation (autodiff) to compute the gradients of scene properties through backpropagation with respect to the light transport simulation. This process is also called differentiable rendering. In this paper, we demonstrate that inverse rendering of plasma properties through differentiable rendering offers an alternative method for obtaining estimates of the underlying physical quantities such as the neutrals density, electron density and electron temperature.

To estimate the physical quantities that influence the light emission and transport, we present a differentiable rendering pipeline where we leverage Null-Scattering~\cite{Miller2019} to simulate light transport in a tokamak, and Path-Replay Backpropagation~\cite{Vicini2021PathReplay} for computing the gradients. Our differentiable plasma rendering pipeline enables not only physically accurate simulation of the imaging system, but also efficient optimisation of the parameters underlying the forward simulation via non-linear optimisation.

We demonstrate that, through our rendering framework, filtered and calibrated high-speed imaging systems in the MAST and MAST-U tokamaks could be used to estimate the full poloidal distributions of electron density, electron temperature and neutrals density. While we demonstrate our method on a simplified representation of the physical distributions, a fully differentiable simulation of the plasma dynamics can exploit the gradients in our pipeline to optimize poloidal and toroidal magnetic fields, neutrals density injection and any other parameters that control the simulation itself. Furthermore, given that there currently exists no diagnostic that can provide full poloidal measurements of electron density and electron temperature, nor spatial measurements of neutral density of any kind, we believe the case-study demonstrated in this paper can pave the way for a new type of diagnostic.

\begin{figure*}
    \centering
    \begin{subfigure}[b]{0.3\textwidth}
        \includegraphics[width=\textwidth]{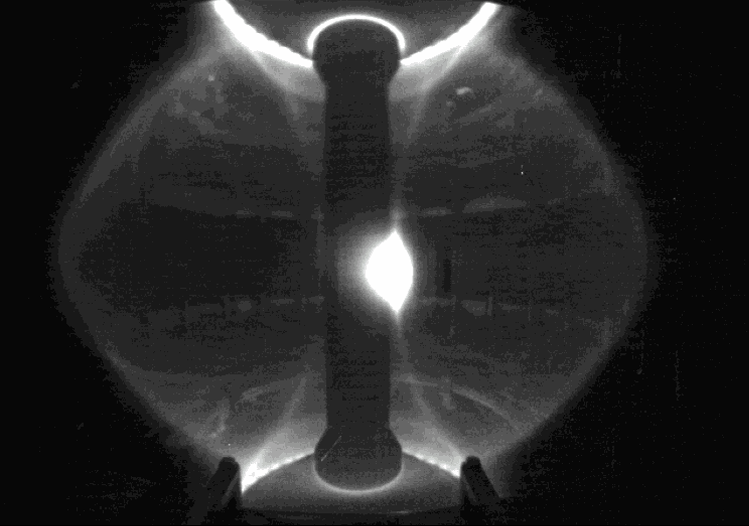}
        \caption{30306 at $0.240\mathrm{s}$}
        \label{subfig:30306-0.240}
    \end{subfigure}
    \hfill
    \begin{subfigure}[b]{0.3\textwidth}
        \includegraphics[width=\textwidth]{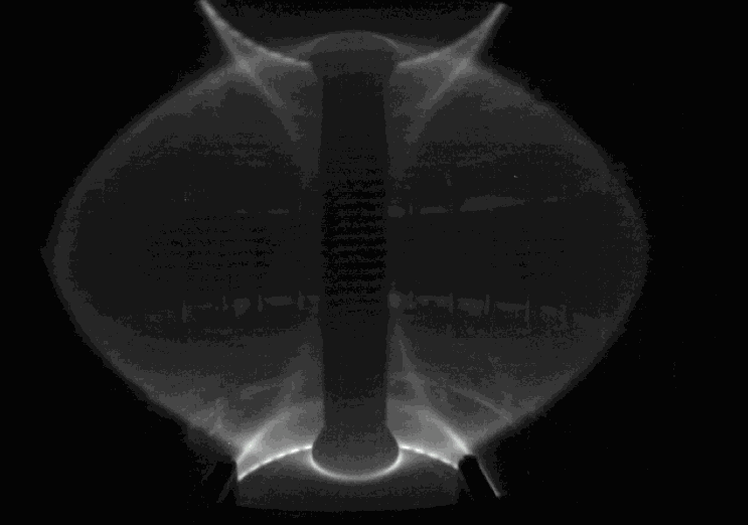}
        \caption{30356 at $0.215\mathrm{s}$}
        \label{subfig:30356-0.215}
    \end{subfigure}
    \hfill
    \begin{subfigure}[b]{0.3\textwidth}
        \includegraphics[width=\textwidth]{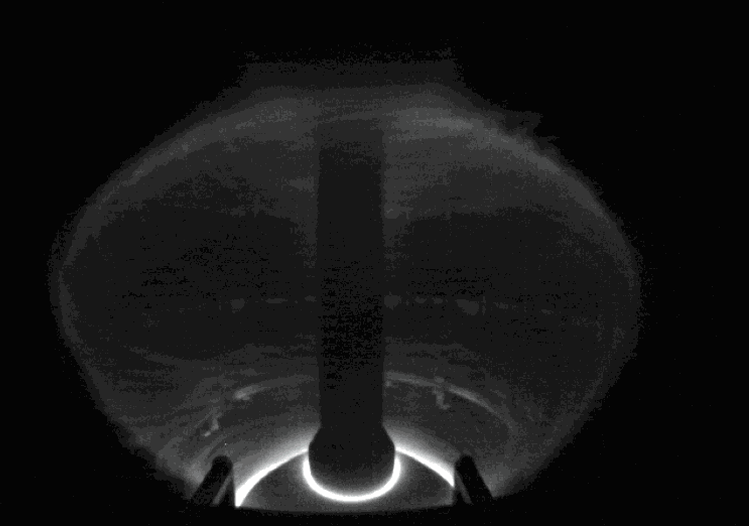}
        \caption{30419 at $0.225\mathrm{s}$}
        \label{subfig:30419-0.225}
    \end{subfigure}
    \caption{Images from shots 30306, 30356 and 30419 in the MAST tokamak, obtained by the Photron APX-RS (RBB) camera mounted on port HM10. These images show the variety of plasma emission configurations.}
    \label{fig:mast-imaging-example}
\end{figure*}

Previous works on reconstructing plasma properties from imaging relied on automated segmentation or hand-labelling and then inferring the properties based on known physical quantities such as the magnetic field in the vessel\cite{10.1063/1.5109470}. The closest to our work is \cite{10.1063/1.5031087} where the emission of a toroidally symmetric plasma is reconstructed from bolometric measurements in the MAST vessel using the Simultaneous Algebraic Reconstruction Technique (SART) and toroidal Green's functions as the emission bases. In the graphics literature, the closest to our method belongs to \cite{10.1145/3528223.3530073}, where synthetic objects are reconstructed by first optimising an emission grid and then using that to infer the scattering albedo of the volume. In this work we demonstrate that similar techniques used in \cite{10.1145/3528223.3530073} can be applied to fusion plasma physics using an appropriate emission model based on Photon Emissivity Coefficients (PEC)~\cite{UniStrathclyde2021}. 

Using a poloidal map representation of the neutrals density, electron density and electron temperature as functions of flux surfaces, our optimisation method converges to a plasma composition that closely reproduces a target camera image, and 1D midplane profiles of the electron and electron temperature. We demonstrate the above principle, on simulation data from a SOLPS run on MAST~\cite{Havlkov2015}. 

In summary, our contributions are:
\begin{itemize}
    \item Differentiable Rendering of fusion plasma,
    \item Framework for optimisation-based diagnostic of global plasma composition,
    \item A low-parameter plasma representation following the flux surfaces.
\end{itemize}

\section{\label{sec:method}Methodology}

The goal of this study is to demonstrate the utility of differentiable simulations for the inference of physical properties through a case-study on the reconstruction of the distributions of the neutrals density ($\neutrals$), electron density ($\electrons$) and electron temperature ($\etemperature$) within the MAST vessel using camera imaging and sparse measurements. Achieving this reconstruction requires physically accurate simulation of the camera imaging diagnostic in MAST, estimation of the gradients of this image with respect to the parameters of the plasma and emission model, and an algorithm that can optimise these parameters to closely reconstruct the reference image.

In the following sections, we go through the algorithmic requirements and implementation of the simulation of light emission and transport, the gradient estimation of this simulation, and define the objective function and the optimisation problem for the reconstruction of the reference image.

\subsection{\label{subsec:inverserendering}The Light Emission and Transport Problem}

The accurate simulation of the camera diagnostic can be split into two parts: Light Emission and Light Transport. The model used for estimating the plasma emission strongly influences the first-order radiance incident on the camera sensor while the light transport method determines the higher-order reflections.

\subsubsection{\label{subsubsec:lightemission}Light Emission}

The plasma emission is determined by three non-equilibrium processes of excitation, recombination and charge-exchange which can be related to the total emission through the Photon Emissivity Coefficients (PECs) obtained from the Open-ADAS library~\cite{UniStrathclyde2021}. The relationship between these three processes and the distributions of neutrals, electrons and temperature is given by 
\begin{equation} \label{eq:pec-full}
    \renewcommand\arraystretch{2.0}
    \begin{array}{rcl}
        \epsilon_{i\rightarrow j} & = & \sum\limits_{\sigma}{\PEC{\sigma}{i}{j}{(exc)}\electrons N_\sigma^{z+}} \\
        & & + \sum\limits_{\rho}{\PEC{\rho}{i}{j}{(rec)}\electrons N_\sigma^{(z+1)+}} \\ 
        & & + \sum\limits_{\rho}{\PEC{\rho}{i}{j}{(exc)}N_HN_\sigma^{(z+1)+}} 
    \end{array}
\end{equation} where $\electrons$ is the electron density, $N_\sigma^{z+}$ is the density of species $\sigma$ in the ionisation state $z+$, and $N_H$ is the density of thermally neutral hydrogen.

For the purposes of this paper, we assume that, for a suitably filtered camera system, the emission from non-Deuterium sources can be ignored, and thus we can further simplify this equation. Assuming $n_{D^+}\approx \electrons$ imposed by the quasi-neutrality constraint, the fact that Deuterium has only one excited state, and that the PECs for charge-exchange are zero, \ref{eq:pec-full} can be simplified to 
\begin{equation} \label{eq:pec-simple}
    \renewcommand\arraystretch{2.5}
    \begin{array}{rcl}
        \epsilon_{i\rightarrow j} & = & \PEC{D}{0}{1}{(exc)}\electrons N_D^0 + \PEC{D}{1}{0}{(rec)}\electrons N_D^{1+} \\
        & = & \PEC{D}{0}{1}{(exc)}\electrons N_D^0 + \PEC{D}{1}{0}{(rec)}\electrons^2
    \end{array}.
\end{equation}

While the PECs are defined for multiple spectral lines, we focus on D-alpha emission in line with a filtered camera system, though our pipeline permits incorporating additional spectral lines by simply summing the emission at each wavelength. Given that there is only a single spectral line, we will denote the total emission as $\emission$.

The PECs are provided in tabulated form as a function of electron density and electron temperature, thus, in order to optimise the density and temperature distributions through the emission, we interpolate the PECs using a locally differentiable bilinear interpolation scheme.

\subsubsection{\label{subsubsec:lighttransport}Light Transport}

Calculating the final image observed by the camera requires simulating the propagation of light within the vessel. This is the light transport problem for which we need an efficient method of tracing light paths that terminate at the camera sensor and originate at our light source, and summing them together such that we have an unbiased estimate. This can be solved via path-tracing methods which are Monte-Carlo methods for constructing light paths, starting from the sensor, and connecting them to the light source. Furthermore, as our light source is volumetric, we need to estimate the emission accrued along paths within a medium. And finally, this method must easily accommodate the calculation of gradients, ideally in an unbiased fashion as well.

Null-Scattering~\cite{Miller2019} is an ideal choice for estimating the light transport inside the plasma medium as it is unbiased and, unlike quadrature methods, scales efficiently with higher-order reflections. This is a technique that, inspired by Woodcock tracking~\cite{woodcock1965techniques} for the paths taken by neutrons in a dense medium, uses a homogenization of the medium and a simple probabilistic view of light interactions to compute the attenuation and transport of light. Considering a non-scattering spatially varying medium where the optical density is $\opticaldensity$, and the emission is $\epsilon$, null-scattering introduces a null density $\nulldensity$ that homogenizes the medium such that $\majorantdensity = \opticaldensity + \nulldensity$ is constant throughout. This homogenized density is then used to sample free-flight distances into the medium using the Beer-Lambert law, $I_d/I_0=\exp{-d\majorantdensity}$ where $d$ is the distance travelled into the medium and selecting, at each of these distances into the medium, an event with probability $\nulldensity/\majorantdensity$ whether or not the ray terminates. This is akin to simulating a single packet of light, sampling a distance into the medium and probabilistically choosing if it is absorbed. As these rays are traced from the camera back to the light source, whenever the ray interacts with the medium we accumulate the emission at the sampled point attenuated by the path taken by that ray. As we are looking at D-alpha emission and D-D plasmas are optically thin in D-alpha line, we take $\opticaldensity=0$ and therefore $\majorantdensity=\nulldensity$.

Furthermore, as our plasma model has many parameters (e.g. 524\,288 for one cross-section of resolution 512 by 1024) and the interior of the vessel is reflective leading to many interactions, we require a method of computing gradients that is memory efficient and mathematically correct. To that end, we use Path-Replay Backpropagation which leverages the reproducibility of pseudo-random numbers to "replay" the forward light transport and accumulate gradients at each interaction in a reverse-mode differentiation context, and unlike tape-based tracking of operations, has linear memory scaling in the number of parameters and reflections~\cite{Vicini2021PathReplay}.

In the above discussion, we have abstracted how light is transported and mentioned that it is being reflected by the interior surfaces of the vessel. In order to simulate these reflections we require a model of the vessel that describes not only the geometry, but also the way in which these surfaces scatter light. For the modelling of these surface reflections we use a Cook-Torrance Microfacet Bidirectional Reflectance Distribution Function (BRDF)~\cite{10.1145/357290.357293} with a Beckmann Normal Distribution Function (NDF)~\cite{beckmann1963scattering}. We assume that all surfaces have a spatially homogeneous fixed albedo of $0.45$ and a fixed roughness of $0.2$. As for the geometry, we use the MAST CAD model, provided courtesy of the MAST Team, to model the interior surfaces, and we will be making the rendering compatible model available publicly.

To compute the final image, the observed radiance incident on the camera sensor must be converted into an image as output by a real-world camera. While there are many characteristics inherent to this transformation, the most prominent of these in a calibrated camera system is the tonemapping, Region-of-Interest Cropping (ROI) and quantisation which we simulate as part of our rendering setup. We use an exposure and gamma-based tonemapping following
\begin{equation}
    p_{ldr}=\cases{{(p_{hdr}\tau)}^\gamma, & for $p_{hdr}\cdot\tau\leq1$ \\ 1, & for $p_{hdr}\cdot\tau>1$}
    \label{eq:tonemapping-exposure-gamma}
\end{equation} where $p_{hdr}$ is the high-dynamic range pixel value, $p_{ldr}$ is the tonemapped, low dynamic range pixel value, $\tau$ is the exposure time and $\gamma$ is the gamma. We choose $\tau=0.9/\mathrm{Q}(p_{hdr}, 97.5\%)$ where $\mathrm{Q}(p_{hdr}, 97.5\%)$ is the $97.5$-percentile of all the HDR pixel values.

\subsection{\label{subsec:syntheticplasma}Modelling Synthetic Plasma}

Similar to prior work, for the reconstruction of emission, we use toroidal Green's functions as the emission bases~\cite{10.1063/1.5031087} which are equivalent to the pixels of a 2D cross-section rotated toroidally about the central axis of the vessel. For representing the plasma distributions of $\electrons$, $\neutrals$ and $\etemperature$, the same representation, while sensible, does not adequately disambiguate line-of-sight emission around the X-points. In order to mitigate this limitation, we use a physically motivated parametric representation of the plasma composition and shape. 

First we apply a change of variables from $\neutrals$, $\electrons$ and $\etemperature$ to the total density ($\totaldensity\ \left[\mathrm{\densityunits}\right]$), the ionisation fraction ($\ionisationfraction\ \left[\ndunits\right]$) and the electron temperature ($\etemperature\ \left[\temperatureunits\right]$ in order to enforce the inverse proportionality between neutrals and electron density. This is achieved with the forward transformation
\numparts
\begin{eqnarray}
    \totaldensity & = \neutrals + \electrons \\
    \nonumber \\
    \ionisationfraction & = \frac{\electrons}{\neutrals+\electrons}\label{eq:bulk-ionisation-temperature-forward} \\
    \nonumber \\
    \etemperature & = \etemperature \\
\end{eqnarray} and the inverse transformation,
\begin{eqnarray}
    \neutrals & = \left(1-\ionisationfraction\right)n \\
    \electrons & = \ionisationfraction n\label{eq:bulk-ionisation-temperature-reverse} \\
    \etemperature & = \etemperature 
\end{eqnarray}
 between these two variables. 
 
In order enforce positivity of the outputs and better represent the high dynamic range of values, we encode the total density and the electron temperature following the transformation
\begin{equation}
    x\mapsto\log_{2}(x+1)\label{eq:log-transformation}.
\end{equation}
 
 Furthermore, in order to account for local variations in density, we split the total density into a bulk density component ($\bulkdensity$) and residual density component ($\residualdensity$) such that
\begin{eqnarray}
    \log_{2}(\totaldensity+1) & = \bulkdensity + \residualdensity & \\
    \mathrm{C_{residual}}: \mean{\left|\residualdensity^{i,j}\right|}{i,j} & \leq \eta \mean{\left|\totaldensity^{i,j}\right|}{i,j}, & \eta < 1
    \label{eq:bulk-residual-balance}
\end{eqnarray}
\endnumparts where $\eta$ controls the ratio between the magnitudes of the bulk and residual densities and $\mean{}{i,j}$ denotes the mean over the the pixels of the cross-section. It should be noted that neither $\bulkdensity$ not $\residualdensity$ are strictly densities as they are summed in log-space to get $\totaldensity$.

In a tokamak, the plasma shape is largely determined by the topology of the magnetic field, typically described by the so-called \textit{flux surfaces}, contours of the poloidal magnetic potential $\psi(R, Z)$, where $R$ and $Z$ are the poloidal coordinates~\cite{Appel2018}. In order to exploit the shape information inherent in the magnetic field configuration, we model $\bulkdensity$, $\ionisationfraction$ and $\etemperature$ as functions of the plasma shape by defining them as tabulated piece-wise linear functions of the 2D normalised $\psi_n$-map. On the other hand, $\residualdensity$ is modelled as a 2D cross-section to allow for local deviations from the bulk density and decouple it from the magnetic field. Thus our full parametric representation of the maps is 
\begin{eqnarray}
    \bulkdensity & = \bulkdensity(\psi_n(R,Z)) \\
    \residualdensity & = \residualdensity(R,Z) \\
    \totaldensity & = \exp_2{\left[\bulkdensity + \residualdensity\right]} - 1 \\
    \etemperature & = \etemperature(\psi_n(R,Z)) \\
    \ionisationfraction & = \ionisationfraction(\psi_n(R,Z)).
\end{eqnarray}

\subsection{\label{subsec:optimisation-configuration}Optimisation}

We consider the reconstruction as an unconstrained optimisation problem in which our objective function is 
\begin{eqnarray}
    \label{eq:total-objective-function}
    \totalobjective = & \renderingobjweight\renderingobjective+\csobjweight\csobjective+ \nonumber\\
        & \midplaneobjweight\midplaneobjective+\tvaobjweight\tvaobjective+\\ 
        & \penaltyobjweight\penaltyobjective \nonumber
\end{eqnarray} where $\renderingobjective$ is the rendering error of the simulation of plasma emission and reflection, $\csobjective$ is the error of the cross-sections themselves, $\midplaneobjective$ is the error of the values of the cross-sections at the midplane, $\tvaobjective$ is the Total Variation Regularisation, $\penaltyobjective$ is the penalty-objective, and $\renderingobjweight$, $\csobjweight$, $\midplaneobjweight$, $\tvaobjweight$, $\penaltyobjweight$ are the weights of each objective term.

We define the Mean Square Error as 
\begin{equation}
    \mserepr{x}{\hat{x}}=\meansquareerror{x}{\hat{x}}{i}
\end{equation} and the Weighted Mean Square Error as
\begin{equation}
    \weightedmserepr{x}{\hat{x}}{\zeta}=\weightedmse{x}{\hat{x}}{i}{\zeta}
\end{equation} allowing us to express the error terms in the above equation in a simpler form. 

\paragraph{Rendering Error} The rendering error is the mean square error between the reference image ($\textrm{img}_{ref}$) and the predicted image ($\textrm{img}_{pred}$), 
\begin{equation}
    \renderingobjective=\mserepr{\textrm{img}_{ref}}{\textrm{img}_{pred}}
\end{equation} where only the under-exposed (pixels with values $\leq1/255$) and over-exposed (pixels with values $\geq1-1/255$) are ignored when computing the error.

\paragraph{Cross-Section Error} The cross-section error is used for demonstrating the efficacy of the parametric representation and is the mean square error between the reference cross-sections ($\textrm{cs}_{ref}$) and the predicted cross-sections ($\textrm{cs}_{pred}$), 
\begin{equation}
    \csobjective=\mserepr{\textrm{cs}_{ref}}{\textrm{cs}_{pred}}.
\end{equation}

\paragraph{Midplane Error} The midplane error is the mean square error between the reference ($\textrm{mid}_{ref}$) and the predicted ($\textrm{mid}_{pred}$) measurements at the midplane, 
\begin{equation}
    \midplaneobjective=\weightedmserepr{\textrm{mid}_{ref}}{\textrm{mid}_{pred}}{\mathrm{max}(\textrm{mid}_{ref})}
\end{equation} normalized by the maximum value of the reference measurements.

\paragraph{Total Variation Error} The Total Variation error, 
\begin{equation}
    \tvaobjective=\mathrm{TVA}(\textrm{cs}_{pred}),
\end{equation} is computed following~\cite{10.1155/2013/217021} and is used to smooth out very high-frequency ringing artifacts resulting from the noise in the predicted image and gradients

\paragraph{Penalty Objective} The penalty-objective, $\penaltyobjective$, arises from the fact that we have constraints on some of our variables while the Amsgrad optimiser is an unconstrained optimiser. The three constraints we have are $\ionisationfraction\geq0$, $\ionisationfraction\leq1$, and $C_{residual}$ defined in \ref{eq:bulk-residual-balance}. In order to satisfy these constraints, we transform them in to an error term following the penalty-objective method~\cite{Luenberger2016}. 
Briefly, for an optimisation problem where the parameters are denoted by $\vec{x}$ and the inequality constraint $g(\vec{x})\leq0$, we can introduce a functional $P$ that takes a function $g(\vec{x})$ defining a constraint of the form $g(\vec{x})\leq0$ and transforms it into a penalty following
\begin{equation}
    \penaltyfunctional{g} = \cases{0, & for $g(\vec{x})\leq0$ \\ g(\vec{x})^2, & for $g(\vec{x})>0$}.
\end{equation} 

Defining the functions $g$ for each of our constraints, we can express them as
\begin{eqnarray}
    \ionisationfraction\geq0 & \mapsto & g_{\ionisationfraction0}(\ionisationfraction) = -\ionisationfraction \\
    \ionisationfraction\leq1 & \mapsto & g_{\ionisationfraction1}(\ionisationfraction) = \ionisationfraction-1 \\
    \mathrm{C_{residual}} & \mapsto & g_{residual} = \mean{\left|\residualdensity^{i,j}\right|}{i,j} - \eta \mean{\left|\totaldensity^{i,j}\right|}{i,j}
\end{eqnarray} with $\mathrm{C_{residual}}$ defined in \ref{eq:bulk-residual-balance}. With this, we can then compute one penalty term for each of our constraints such that
\begin{equation}
    \penaltyobjective = \penaltyfunctional{g_{\ionisationfraction0}}(\ionisationfraction) + \penaltyfunctional{g_{\ionisationfraction1}}(\ionisationfraction) + \penaltyfunctional{g_{residual}}(\totaldensity, \residualdensity).
\end{equation}

\section{\label{sec:results}Results}

\subsection{\label{subsec:experimental-setup}Experimental Setup}

We implement the described differentiable rendering and optimisation pipeline using Mitsuba 3~\cite{Mitsuba3} for the differentiable rendering and PyTorch~\cite{Ansel_PyTorch_2_Faster_2024} for the optimisation. We chose Mitsuba 3 as it can easily run on both CPUs and GPUs, and was the simplest to incorporate the chosen rendering algorithms into. PyTorch has a more mature set of optimisation tools so we opted to use it for most of the mathematical operations and the optimisation algorithms. All of our experiments are run on a single Nvidia A100-SXM4-80GB GPU on the CSD3 HPC Platform.

Given that many different optimisation problems can be considered within the same framework described above, we consider 3 different experiments of increasing difficulty where we look at reconstructing
\begin{enumerate}
    \item \label{experiment:neutrals-only} the $\neutrals$ distribution given known $\electrons$ and $\etemperature$ distributions, demonstrated in~\ref{subsec:recovering-neutrals-only},
    \item \label{experiment:all-maps-full-res} the distributions of $\neutrals$, $\electrons$, and $\etemperature$, demonstrated in~\ref{subsec:recovering-all-maps-full-sensor},
    \item \label{experiment:all-maps-cropped-quantised} the distributions with an ROI cropped sensor image and quantisation of the image, demonstrated in~\ref{subsec:recovering-all-maps-roi-quantised},
\end{enumerate}

As diagnostic data from experiments does not capture measurements at every spatial point in the vessel, and is subject to noise and requires calibration, we use a known SOLPS simulation case on MAST shot 30356~\cite{Havlkov2015} to validate our framework. As the SOLPS simulation uses a mesh with spatially varying resolution, we bi-linearly interpolate these values onto fixed resolution maps with a resolution of 512 by 1024. We use a segmentation map to identify the pixels in the fixed resolution maps that correspond to the SOLPS mesh, and render and optimise only the pixels corresponding to the mesh as we do not have any priors for regions outside the mesh.

\begin{figure*}
    \centering
    \includegraphics[width=0.95\textwidth]{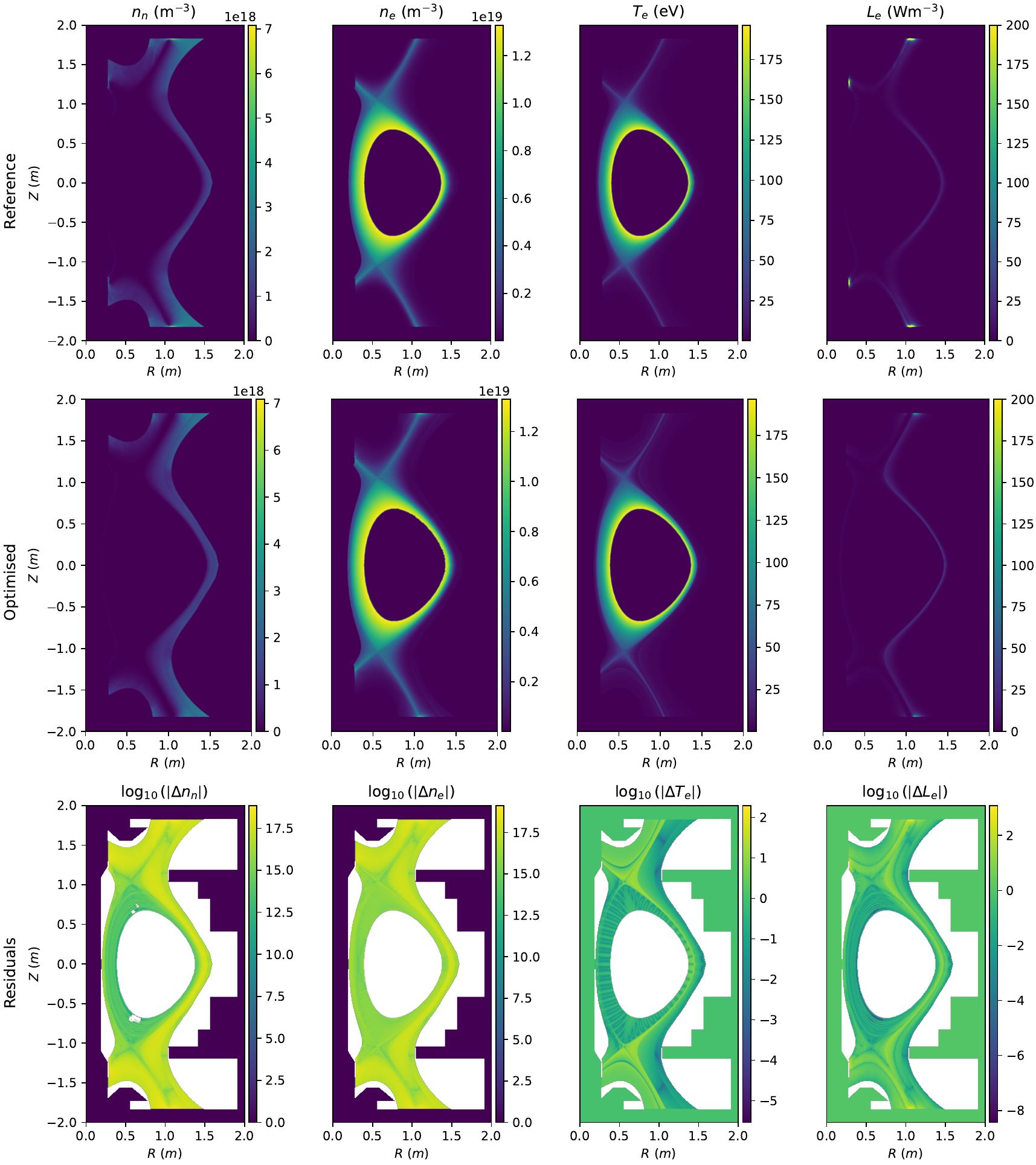}
    \caption{First Row: Cross-sections obtained from a SOLPS simulation of MAST shot 30356~\cite{Havlkov2015} as interpolated onto a regular 2D grid. Second Row: Parametric representation of the cross-sections using the formulation described in~\ref{subsec:syntheticplasma}. Third Row: Residuals of the parametric reconstruction. The absolute $\log_{\mathrm{10}}$ difference between the maps in the first row and the second row.}
    \label{fig:solps-example-sections-with-parametric}
\end{figure*}

The first row in Figure~\ref{fig:solps-example-sections-with-parametric} shows the cross-sections obtained from the SOLPS simulation and we note that this simulation extends only partially into the plasma core. While we expect the emission in the core to be low in actual experiments due to the sparsity of neutrals, any emission from this region would be included in high-speed camera images and may provide insight into the composition in the core. Similarly, Thomson Scattering measurements also include measurements from the core which we lack in our mid-plane profiles as they are absent from the SOLPS mesh. The second row shows the capability of our parametric representation and is a validation that this is a useful way of modelling synthetic plasma.

\begin{figure*}
    \centering
    \begin{subfigure}[t]{0.475\textwidth}
        \includegraphics[width=\textwidth]{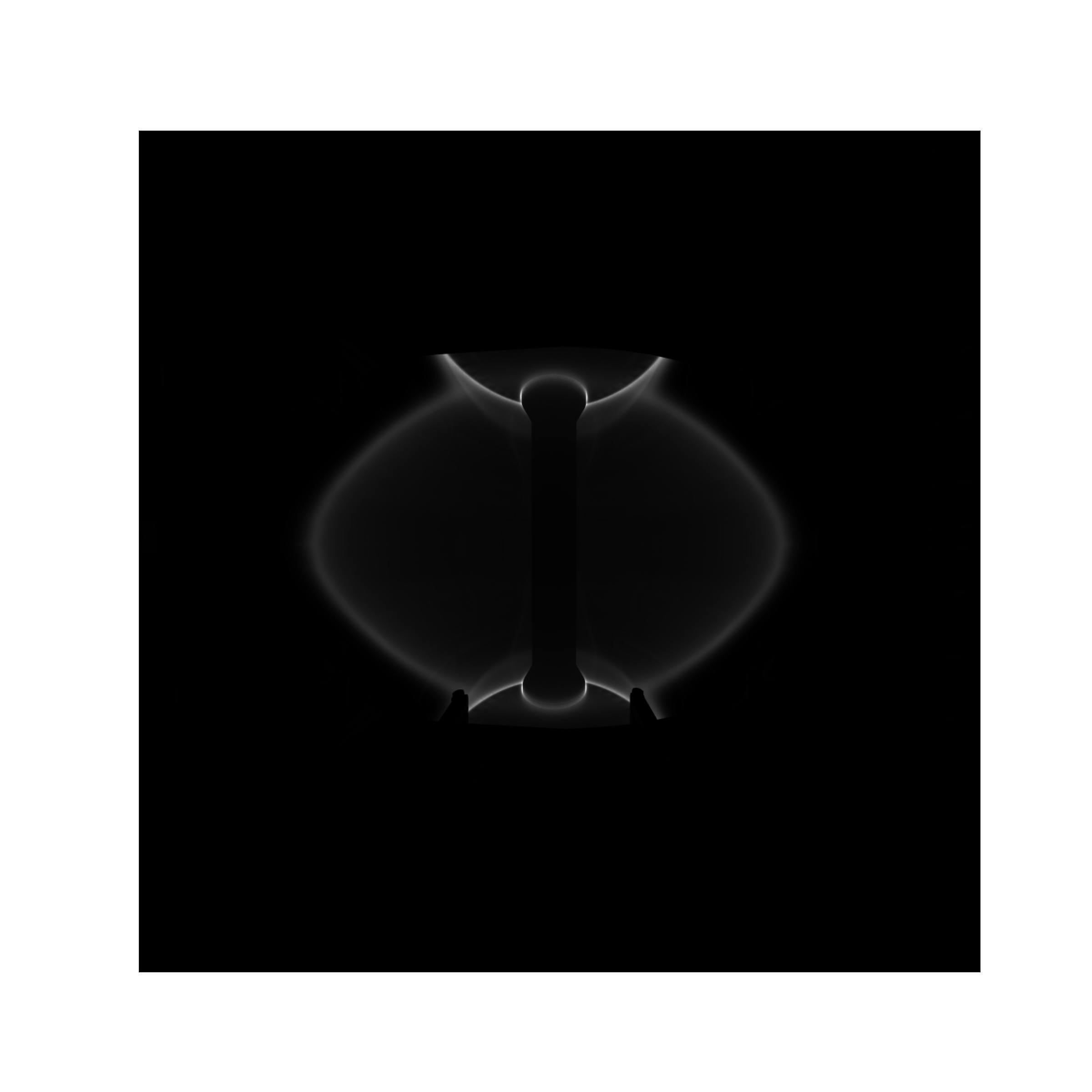}
        \caption{High-dynamic range render scaled to range $\left[0,1\right]$.}
        \label{subfig:solps-hdr-render}
    \end{subfigure}
    \hfill
    \begin{subfigure}[t]{0.475\textwidth}
        \includegraphics[width=\textwidth]{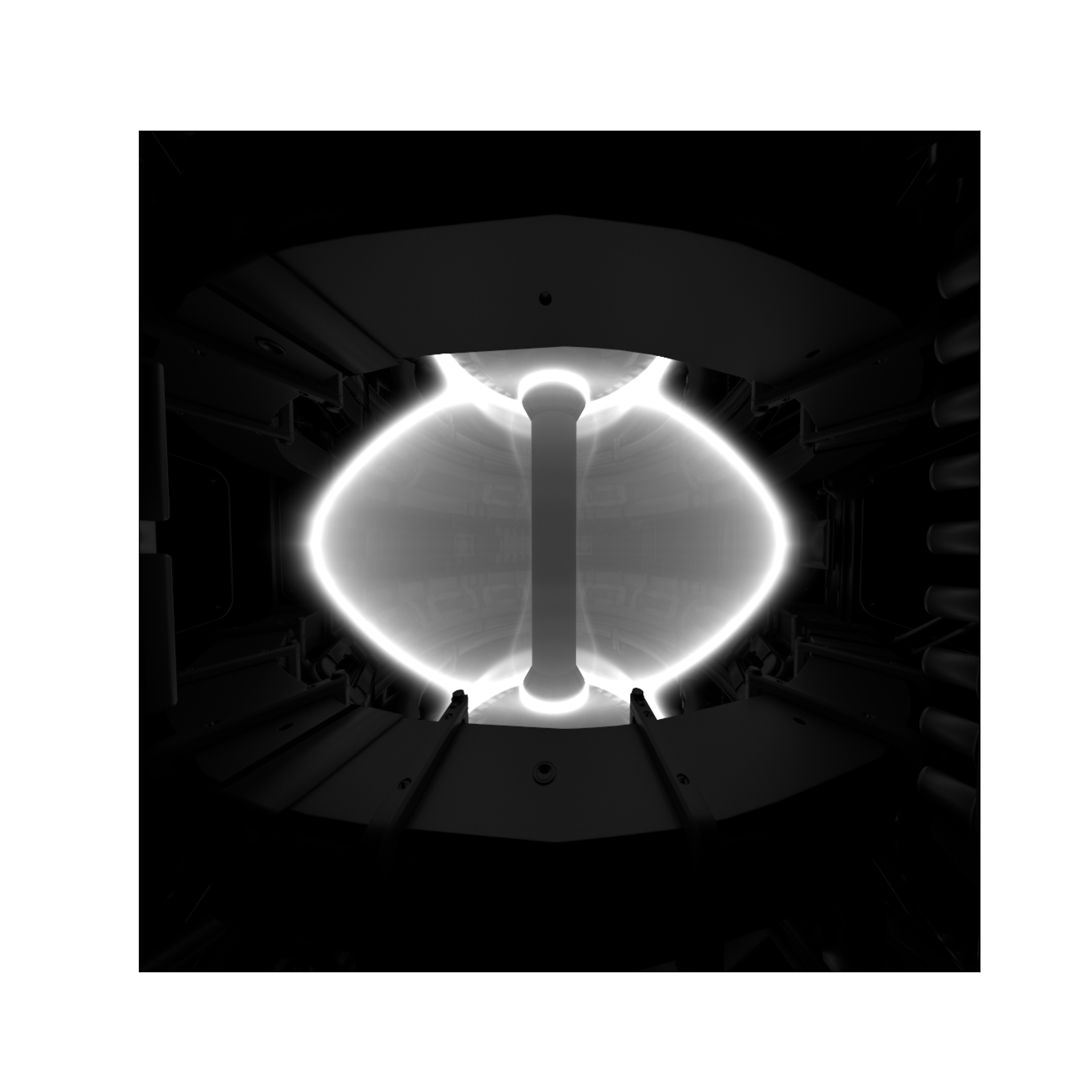}
        \caption{Tonemapped render using an exposure time of $0.109\mathrm{s}$ and a gamma of $0.625$.}
        \label{subfig:solps-ldr-render}
    \end{subfigure}
    \caption{Rendered using Mitsuba 3~\cite{Mitsuba3} from SOLPS cross-sections using PEC coefficients obtained from Open-ADAS~\cite{UniStrathclyde2021} to compute the volumetric emission.}
    \label{fig:solps-render}
\end{figure*}

Figure~\ref{fig:solps-render} shows the rendering of the plasma parameters from SOLPS using $1024$ samples per pixel which takes approximately 6 minutes to render. The camera position and focal length has been calibrated using the CALCAM software~\cite{calcam} and the MAST CAD model, following the same procedure as used in~\cite{10.1063/1.5109470} and the parameters of which are in~\ref{appendix:calibration-parameters}. Using the tonemapping in~\ref{eq:tonemapping-exposure-gamma}, we set the exposure to $0.109\mathrm{s}$ for which ensures that most of the pixels are well-exposed, and the gamma to $0.625$ as this qualitatively matches the real images seen in Figure~\ref{fig:mast-imaging-example}. For Experiments~\ref{experiment:neutrals-only} and \ref{experiment:all-maps-full-res}, the exposure is set to $0.109\mathrm{s}$ and for Experiment~\ref{experiment:all-maps-cropped-quantised}, the exposure is set to $0.0794\mathrm{s}$ as there are proportionally more bright pixels than dark.

Qualitatively, these renders closely resemble the images shown in Figure~\ref{fig:mast-imaging-example} in terms of the overall distribution of intensities, the appearance of reflections, and the primary regions of emission. They differ in that the high-speed camera used to capture the images in Figure~\ref{fig:mast-imaging-example} is also subject to systematic and random noise which our rendering pipeline excludes.

\subsection{\label{subsec:recon-cross-sections}Reconstructing Cross-Sections}

As our reference cross-sections have a resolution of 512 by 1024 in (R,Z)-poloidal coordinates, and while reconstructing the cross-sections could be done directly with 512 by 1024 maps and high-resolution tabulated functions, the non-linearity of the optimisation makes this a harder problem due to the high dimensionality. To accelerate the optimisation and reduce the complexity, we use a hierarchical optimisation approach where we, at each iteration, solve a lower dimensional problem and then upscale it into higher dimensions until we achieve the final result.

For Experiment \ref{experiment:neutrals-only}, we start with a $\neutrals$ cross-section of resolution 32 by 64, up-scaling it by a factor of 4 using bilinear interpolation 3 times until reaching a final resolution of 512 by 1024. We assume an initial uniform density of $\expten{17}$ as this avoids both significant over- and under-exposure.

For Experiments \ref{experiment:all-maps-full-res} and \ref{experiment:all-maps-cropped-quantised}, we start with a $\residualdensity$-map with resolution 32 by 64, and a tabulated function of $\bulkdensity$, $\ionisationfraction$ and $\etemperature$ with 64 segments ranging from $\psi_n=0.769$ to $\psi_n=1.42$ which are the minimum and maximum values of $\psi_n$ on the SOLPS mesh. The $\residualdensity$-map and the tabulated functions are then up-scaled by a factor of 4 using bilinear and linear scaling respectively 3 times to achieve a final resolution of 512 by 1024 and 1024 segments respectively. Each of our maps are initialized as follows:
\begin{enumerate}
    \item the $\residualdensity$-map is initialised to a value of $1$ to avoid zero gradients, 
    \item the $\bulkdensity$-function is initialised to a linearly decaying value corresponding to $\bulkdensity=20$ at $\psi_n=0$ to $\bulkdensity=19$ at $\psi_n=1$, 
    \item the $\etemperature$-function to a linearly decaying value corresponding to $\etemperature=2\scinot{4}$ at $\psi_n=0$ to $\etemperature=2\scinot{1}$ at $\psi_n=1$, 
    \item the $\ionisationfraction$-function to a linearly decaying value corresponding to $\ionisationfraction=1$ at $\psi_n=1$ to $\ionisationfraction=0$ at $\psi_n=2$.
\end{enumerate}

\subsection{\label{subsec:optimisation-setup}Optimisation Setup}

The optimisation framework discussed in \ref{subsec:optimisation-configuration} permits many ways of selecting the initial step size, how many optimisation steps to take, how to vary the step size over the course of the optimisation, and when to terminate the optimisation. We use the default step size of Amsgrad, $\expten{-3}$ in Experiments~\ref{experiment:all-maps-full-res} and \ref{experiment:all-maps-cropped-quantised}, and a step size of $1.0$ in Experiment~\ref{experiment:neutrals-only}. We increase the number of optimisation steps at each up-scaling stage, roughly quadrupling the number of optimisation steps each time. We reduce the step size using PyTorch's `ReduceLROnPlateau`~\cite{Ansel_PyTorch_2_Faster_2024} which tracks the improvement of the error and reduces it whenever it does not improve for `patience` number of steps, we use the default values with the patience increased linearly at each up-scaling stage. And we terminate the optimisation whenever reducing the step size by a factor of $\expten{2}$ stops improving the error.

Given that the rendering is subject to noise, $\renderingobjective$ is a stochastic estimate of the true error. In order to account for this, the samples per pixel at each iteration of the optimisation is increased monotonically following $\round{2.25\cdot{\renderingobjective}^{-0.5}}$. This is roughly the variance scaling of the underlying rendering and thus ensures that, as the error decreases, we can resolve the error and the gradients with accuracy proportional to the error itself. This also has the advantage of accelerating the optimisation as we avoid using high samples per pixel in the early stages of the optimisation. Additionally, as the sampling rate of the emission is proportional to $\nulldensity$ and thus increasing this gives a more accurate estimate of the emission for each ray that passes through the medium, we set its value to $\exptwo{\mathrm{u}+2}$ where $u$ is the upscaling stage ranging from 0 to 2 for the 3 upscaling stages.

\subsection{\label{subsec:objective-functions}Objective Functions}

Depending on the reconstruction experiment, we use a different weighting of each of the terms in \ref{eq:total-objective-function}. 

For Experiment \ref{experiment:neutrals-only}, we set $\midplaneobjweight=0$, $\csobjweight=0$ and $\penaltyobjweight=0$ as we do not have mid-plane neutrals measurements, we are attempting to optimise through the rendering itself and so we do not explicitly enforce a match to the reference cross-sections, and the encoding in \ref{eq:log-transformation} explicitly enforces the positivity of the cross-section. Furthermore, we do not mask out pixels for this experiment as the under- and over-exposed pixels force the estimated $\neutrals$ in these regions to attain a lower bound and an upper bound respectively of the true neutrals density. As we do not want to overly smooth the optimised cross-section, we set $\tvaobjweight=\expten{-8}$.

For Experiments \ref{experiment:all-maps-full-res} and \ref{experiment:all-maps-cropped-quantised}, we set $\csobjweight=0$ term as, again, we want to reconstruct the cross-sections without explicitly enforcing a match to the reference cross-sections. As we do not want to overly smooth the optimised cross-section, we set $\tvaobjweight=\expten{-3}$.

\subsection{\label{subsec:recovering-neutrals-only}Experiment~\ref{experiment:neutrals-only}: Reconstructing the Neutrals Density}

\begin{figure*}
    \centering
    \includegraphics[width=0.95\textwidth]{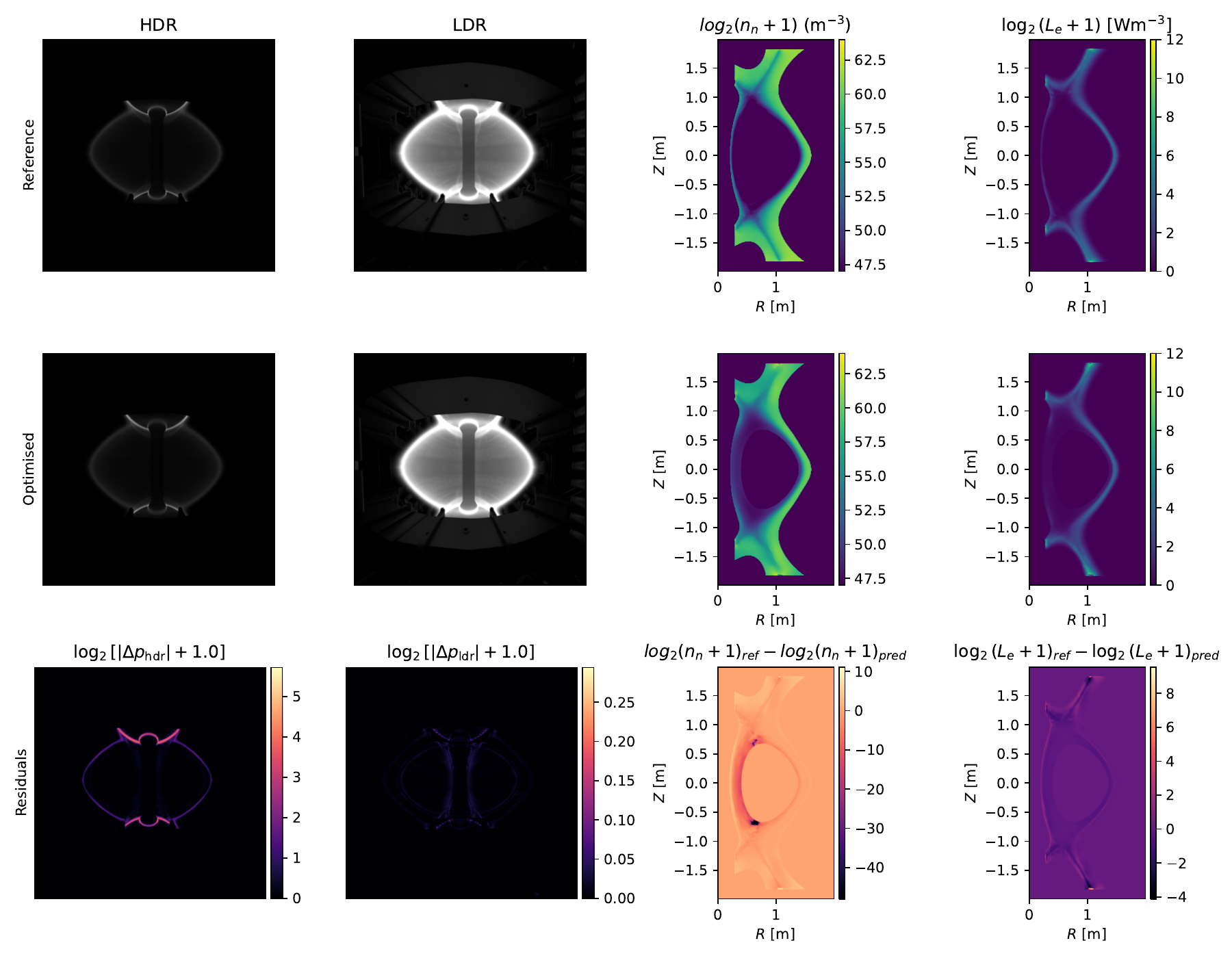}
    \caption{Experiment~\ref{experiment:neutrals-only}: Neutrals reconstruction using $\electrons$ and $\etemperature$ as known priors. The first row shows the reference images and cross-sections, the second row shows the reconstructions and the last row shows the residuals of these reconstructions. The renders are tonemapped using $\tau=0.109\mathrm{s}$ and $\gamma=0.625$.}
    \label{fig:neutrals-recon}
\end{figure*}

The neutrals density reconstruction, shown in Figure~\ref{fig:neutrals-recon}, achieves a rendering error of $1.58\scinot{-5}$ and a mean square error on the log-encoded cross-section, $\log_2{(\neutrals+1)}$, of $5.97$ compared to an error of $1.95\scinot{1}$ for the initial value of $\neutrals=\expten{17}$. The mean square error on the underlying $\neutrals$ map is $1.07\scinot{35}$

We see that the differentiable rendering routine achieves a plausible reconstruction of the neutrals density where the core neutrals is low and the density in the rest of the vessel is reasonably distributed. In comparison to the reference cross-section, this reconstruction captures the fine features of the legs, the location of the LCFS and the distribution of densities around the X-Points. It is also able to accurately reproduce the locally high-emission at the divertor strikes that results from the reflection of neutrals from the vessel walls.

\subsection{\label{subsec:recovering-all-maps-full-sensor}Experiment~\ref{experiment:all-maps-full-res}: Reconstructing $\neutrals$, $\electrons$, and $\etemperature$ using the Full-Sensor}

\begin{figure*}
    \centering
    \includegraphics[width=0.95\textwidth]{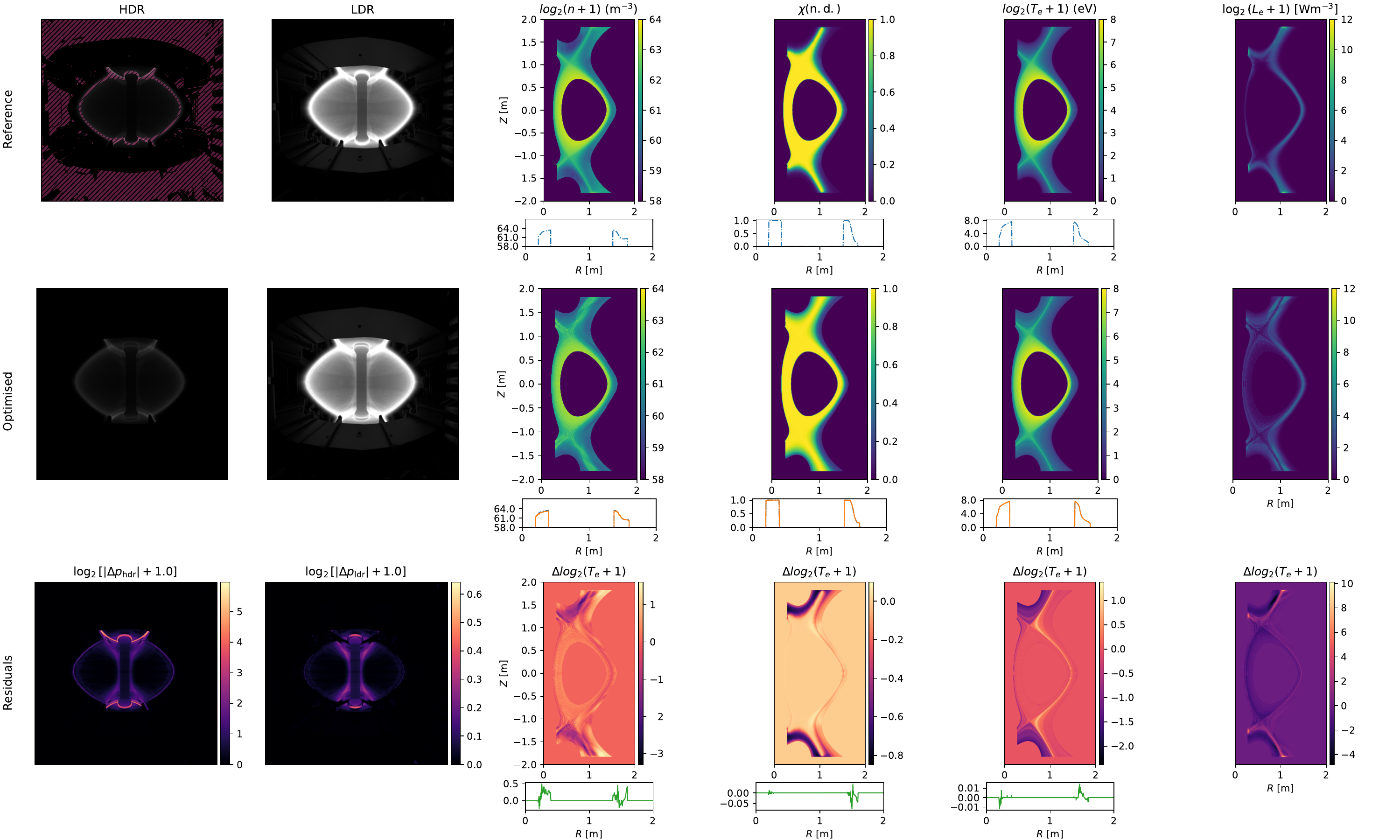}
    \caption{Experiment~\ref{experiment:all-maps-full-res}: Reconstruction of total density $n$, ionisation fraction $\ionisationfraction$ and electron temperature $\etemperature$ using renders and midplane values as the ground truth. The first row shows the reference images and cross-sections while the second row shows the reconstructions. The renders are tonemapped using $\tau=0.109\mathrm{s}$ and $\gamma=0.625$.}
    \label{fig:recon-indirect}
\end{figure*}

In Figure~\ref{fig:recon-indirect}, we show the results of reconstructing the distributions of $\neutrals$, $\electrons$, and $\etemperature$ given the full sensor image. This optimisation took roughly $26$ hours and we see that it is able to achieve a mean square rendering error of $8.30\scinot{-4}$ on the tonemapped, low dynamic range image, and a mean square error of $1.53$ on the high dynamic range image. This translates to a very close match to the ground truth low dynamics range image, but, as can be seen at the divertor strike points, differs significantly due to mismatch in intensity. Part of this mismatch is due to the fact that we filter out pixels that are over-exposed and thus the optimisation receives no signal from these pixels, but another part is due to the fact that these pixels are clamped to a value of $1$ despite being significantly brighter leading to the higher error seen in the high dynamic range image.

We see that not only does the render match the ground truth image, but the underlying maps also accurately reconstruct the ground truth maps. The midplane profiles of all three maps closely match the ground truth and we see that this is translated well off-midplane. Furthermore, the density is well reconstructed across the entire map, with the expected shape and capturing some of the local variation in density. While there are still some high frequency artifacts along the flux surfaces, this is due to the stochasticity of the optimisation and could be further smoothed out with postprocessing. Furthermore, we report the exact errors achieved on these maps using two different metrics in \ref{appendix:errors-on-cross-sections}.

\subsection{\label{subsec:recovering-all-maps-roi-quantised}Experiment~\ref{experiment:all-maps-cropped-quantised}: Reconstructing with a Cropped and Quantised Image}

\begin{figure*}
    \centering
    \includegraphics[width=0.95\textwidth]{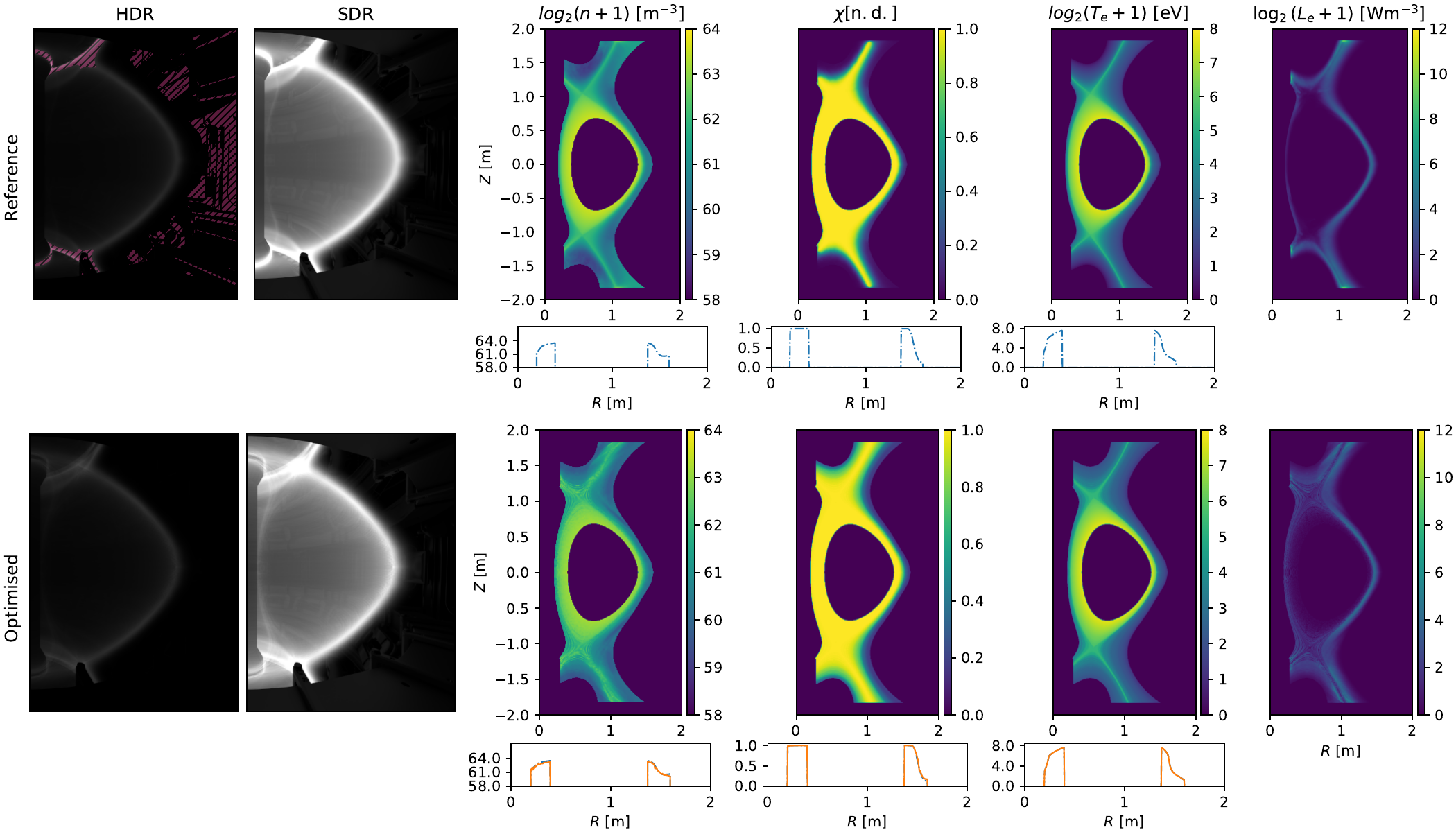}
    \caption{Experiment~\ref{experiment:all-maps-cropped-quantised}: Reconstruction of total density $n$, ionisation fraction $\ionisationfraction$ and electron temperature $\etemperature$ using cropped and quantised renders, and midplane values as the ground truth. The first row shows the reference images and cross-sections while the second row shows the reconstructions. The renders are tonemapped using $\tau=0.0794\mathrm{s}$ and $\gamma=0.625$.}
    \label{fig:recon-indirect-w-roi-quantised}
\end{figure*}

In Figure~\ref{fig:recon-indirect-w-roi-quantised}, we show the results of Experiment~\ref{experiment:all-maps-cropped-quantised} where the render is cropped and the reference image is 8-bit quantised. The rendering mean square error achieved by this optimisation is $1.71\scinot{-3}$ for the low dynamic range image and $3.49$ for the high dynamic range image where we observe a similar discrepancy at the divertor strikes points. 

We see that this optimisation achieves comparable results to the reconstruction with the full sensor despite a) running faster due to the reduced number of pixels in the rendering, and b) the presence of quantisation effects on the pixel values. The biggest difference compared to the full sensor is the increased noise in the density distributions.

\subsection{\label{subsec:robustness}Robustness of the Optimisation to the Initialisation}

In~\ref{subsec:recovering-all-maps-full-sensor} and~\ref{subsec:recovering-all-maps-roi-quantised}, we demonstrated successful reconstruction of the underlying maps starting from a specific initialisation of the maps and tabulated functions. Here we show the robustness of the method with regards to these initialisations, specifically in varying the ionisation fraction, $\ionisationfraction$. 

We choose $\ionisationfraction$ to vary as the emission is proportional to the product $\neutrals\cdot\electrons$ and thus decreasing $\neutrals$ can, locally, be compensated by a proportional increase in $\electrons$. We demonstrate this many-to-one mapping issue using the same setup as the ROI cropped and quantised optimisation, except we set $\ionisationfraction$ to $0.5$ at the extremes of the SOLPS mesh instead of $0$.
 
\begin{figure*}
    \centering
    \includegraphics[width=0.95\textwidth]{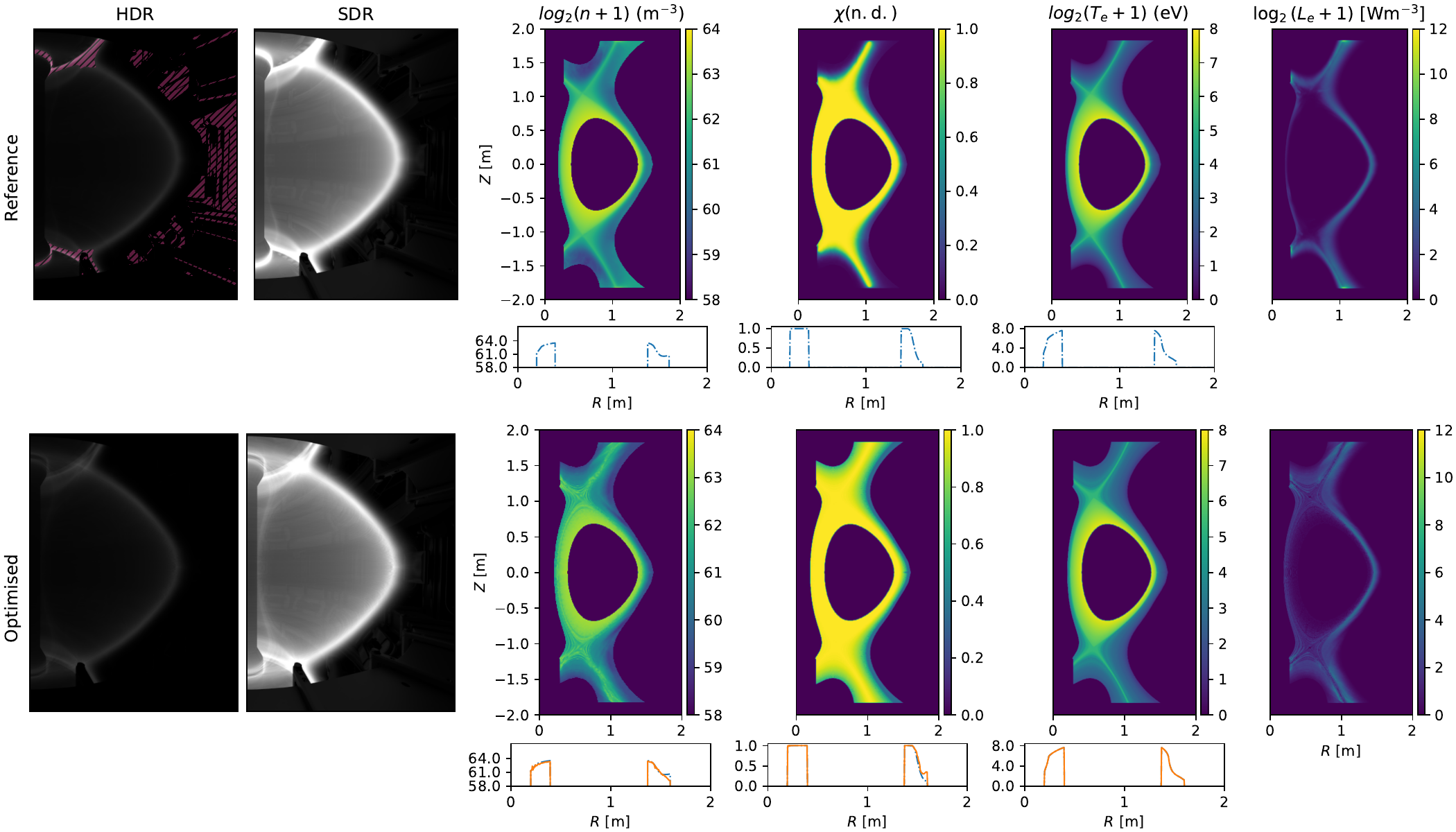}
    \caption{Reconstruction following same setup as Experiment~\ref{experiment:all-maps-cropped-quantised} with the ionisation fraction initialised to $0.5$ at the mesh boundary. The first row shows the reference images and cross-sections while the second row shows the reconstructions. The renders are tonemapped using $\tau=0.0794\mathrm{s}$ and $\gamma=0.625$.}
    \label{fig:recon-indirect-w-roi-quantised-diff-init}
\end{figure*}

We find that with this initialisation, as shown in Figure \ref{fig:recon-indirect-w-roi-quantised-diff-init}, the optimisation is able to minimise the render and midplane errors. In comparison to Experiment~\ref{experiment:all-maps-cropped-quantised}, we achieve a mean square error of $1.31\scinot{-3}$ on the tonemapped render, and $3.50$ on the HDR render. In the regions where there is no signal in the image, it is unable to discern the true distribution of neutrals density, and under-estimates the neutrals distribution outside of the LCFS as indicated by the higher than ground truth ionisation fraction.

\section{\label{sec:conclusion}Conclusion}

In this paper, we have demonstrated that differentiable rendering can serve as a valuable tool for inverse estimation of plasma parameters from image data. While previous works exist in the form of rendering and least-squares fitting to image measurements, our technique enables end-to-end rendering and reconstruction with better computational scaling, and easy integration of multi-modal measurements in the form of midplane measurements.

Our contributions include the reconstruction of the neutrals density distribution which, to our knowledge, almost always required expensive Monte-Carlo simulations to reconstruct prior to this technique. As our method recovers plausible distributions of neutrals density from imaging and measurements, our reconstructions can complement SOLPS simulations with neutrals transport by providing validation and comparison to experiments, and perhaps even providing a better prior for the simulations at a relatively modest computational cost. While we have reported timings of the optimisation and rendering on a single GPU, the algorithm itself scales trivially and thus could be parallelised easily across multiple GPUs leading to further speed-ups in the process.

The reconstruction pipeline demonstrated in our work also motivates the use of calibrated and filtered imaging systems on the MAST-U tokamak in the near future as this would enable new 2D diagnostic mapping of various plasma quantities, including electron density and temperature distributions, neutrals distribution, and eventually impurity distributions. Furthermore, as our rendering pipeline is agnostic to how the emission is computed, additional diagnostics such as bolometry, and physical models such as simulations themselves can be used to constrain the space of possible distributions.

As demonstrated in~\ref{subsec:robustness}, due to the difficulty of inverting many-to-one mappings, we expect that future work will include statistical modelling of likely inversions based on the input image and measurements. A possible direction is to train a neural network on a dataset of pairs of images and midplane measurements, and estimate the underlying distributions of $\neutrals$, $\electrons$, and $\etemperature$.

We have also shown that the availability of gradients greatly improves the convergence and stability of non-linear optimisation based inverse estimation. As auto-differentiation is not only limited to light transport simulations, we hope that our work will encourage the adoption of differentiable codes in the wider physics community.

\section*{Acknowledgements}

We would like to thank David Moulton for sharing with us both the SOLPS data and the code for reading it. This work was partly supported by the UK Atomic Energy Authority (Agreement Number 12038). Pieter Peers was supported in part by NSF grant IIS-1909028 and DOE grant DE-SC0024624. This work has been carried out within the framework of the EUROfusion Consortium, funded by the European Union via the Euratom Research and Training Programme (Grant Agreement No 101052200 — EUROfusion) and from the EPSRC [grant number EP/T012250/1].  To obtain further information on the data and models underlying this paper please contact PublicationsManager@ukaea.uk*. Views and opinions expressed are however those of the author(s) only and do not necessarily reflect those of the European Union or the European Commission. Neither the European Union nor the European Commission can be held responsible for them. This work was performed using resources provided by the Cambridge Service for Data Driven Discovery (CSD3) operated by the University of Cambridge Research Computing Service (www.csd3.cam.ac.uk), provided by Dell EMC and Intel using Tier-2 funding from the Engineering and Physical Sciences Research Council (capital grant EP/T022159/1), and DiRAC funding from the Science and Technology Facilities Council (www.dirac.ac.uk).

\section*{References}

\bibliographystyle{iopart-num}
\bibliography{main_references}

\appendix

\section{\label{appendix:calibration-parameters}Camera Calibration}

\begin{figure*}
    \centering
    \includegraphics[width=0.95\textwidth]{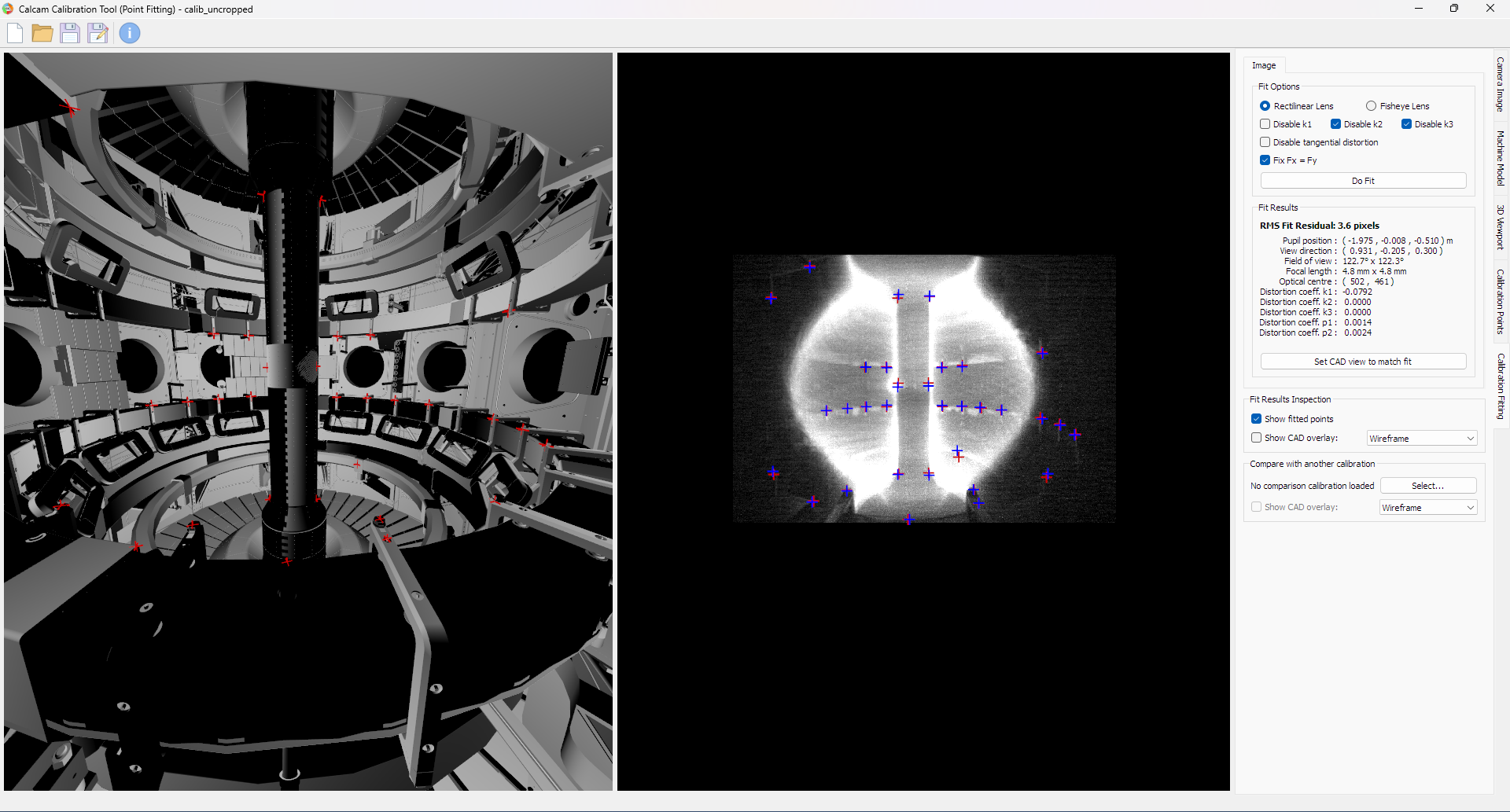}
    \caption{Camera calibration showing the MAST CAD alongside a frame from shot 30301 in CALCAM. The camera image has been brightened and the contrast decreased in order to bring out the visibility of the vessel interior.}
    \label{fig:camera-calibration}
\end{figure*}

The camera calibration is done through the use of the MAST CAD model and the CALCAM software. In order to calibrate the image and correctly estimate the distortion parameters, we first pad the incoming image with zeros on each side based on the sensor pixel width/height and the Region-of-Interest (ROI) crop that was used. Without this step the camera position and orientation would be incorrectly estimated in relation to a synthetic camera using the same ROI. This also enables us to simulate different ROI crops without having to recalibrate the camera position and orientation. Figure \ref{fig:camera-calibration} shows a screenshot from the CALCAM software showing the number of points used for calibration and Eq. \ref{eq:calibrated-parameters} shows the calibrated camera parameters.

\begin{eqnarray}
\label{eq:calibrated-parameters}
    \left(p_x,p_y,p_z\right)&=\left(-1.975,-0.008,-0.510\right)\mathrm{m} \\
    \left(d_x,d_y,d_z\right)&=\left(0.931,-0.205,0.300\right)\mathrm{m} \\
    \left(f_x,f_y\right)&=\left(4.8,4.8\right)\mathrm{mm} \\
    \left(o_x,o_y\right)&=\left(502,461\right)\mathrm{px} \\
    \left(k_1,k_2,k_3\right)&=\left(-0.0792,0.0,0.0\right) \\
    \left(p_1,p_2\right)&=\left(0.0014,0.0024\right) \\
\end{eqnarray}

In this paper we do not utilise the distortion parameters nor the offset optical center, and are only interested in the camera extrinsics and the focal length. We also note that the camera moves during and between shots and therefore these calibrations would be updated for optimisation on real images.

\section{\label{appendix:errors-on-cross-sections}Cross-Sections Errors from Experiments}

Here we report the exact errors achieved by the experiments on the various underlying maps and representations. Our metrics are the mean square error and the mean absolute error. In~\ref{table:recon-errors-full-sensor} we show the errors achieved on Experiment~\ref{experiment:all-maps-full-res}, the full sensor reconstruction, in~\ref{table:recon-errors-roi-crop} we show the errors achieved on Experiment~\ref{experiment:all-maps-cropped-quantised}, the ROI cropped and quantised image reconstruction, and in~\ref{table:recon-errors-roi-crop-diff-init} we show the errors achieved on the modified Experiment~\ref{experiment:all-maps-cropped-quantised} where the ionisation fraction, $\ionisationfraction$, is initialized to a higher value of $0.5$ on the SOLPS mesh boundary.

\begin{table*}
    \centering
    \caption{Errors achieved by the optimisation of the parametric maps using the renders and midplane measurements, both in the parametric representation and in the SOLPS representation. MSE is the Mean Square Error and MAE is the Mean Absolute Error.}
    \label{table:recon-errors-full-sensor}
    \renewcommand{\arraystretch}{1.5}
    \begin{tabular}{c||c|c||c|c}
        & \multicolumn{4}{c}{Errors on Experiment~\ref{experiment:all-maps-full-res}} \\
        \cline{2-5}
        &  \multicolumn{2}{c}{MSE} & \multicolumn{2}{c}{MAE} \\
        \cline{2-5}
        & Whole Map & Midplane & Whole Map & Midplane \\
        \hline
       $\log_2{(\totaldensity+1)}$ & $8.89\scinot{-2}$ & $1.00\scinot{-2}$ & $1.23\scinot{-1}$ & $3.83\scinot{-2}$ \\
       $\ionisationfraction$       & $1.69\scinot{-2}$ & $5.62\scinot{-5}$ & $3.63\scinot{-2}$ & $1.76\scinot{-3}$ \\
       $\log_2{(\etemperature+1)}$ & $7.92\scinot{-2}$ & $3.24\scinot{-6}$ & $9.17\scinot{-2}$ & $4.74\scinot{-4}$ \\
        \hline
       $\neutrals$     & $9.60\scinot{34}$ & $2.18\scinot{33}$ & $7.94\scinot{16}$ & $9.05\scinot{15}$ \\
       $\electrons$    & $6.71\scinot{35}$ & $3.19\scinot{35}$ & $3.22\scinot{17}$ & $1.94\scinot{17}$ \\
       $\etemperature$ & $6.62\scinot{0} $ & $9.95\scinot{-4}$ & $7.15\scinot{-1}$ & $7.02\scinot{-3}$ \\
       $\emission$     & $4.39\scinot{-1}$ &                   & $2.32\scinot{-1}$ &                   
    \end{tabular}
\end{table*}

\begin{table*}
    \centering
    \caption{Errors achieved by the optimisation of the parametric maps using the renders and midplane measurements, both in the parametric representation and in the SOLPS representation. MSE is the Mean Square Error and MAE is the Mean Absolute Error.}
    \label{table:recon-errors-roi-crop}
    \renewcommand{\arraystretch}{1.5}
    \begin{tabular}{c||c|c||c|c}
        & \multicolumn{4}{c}{Errors on Experiment~\ref{experiment:all-maps-cropped-quantised}} \\
        \cline{2-5}
        &  \multicolumn{2}{c}{MSE} & \multicolumn{2}{c}{MAE} \\
        \cline{2-5}
        & Whole Map & Midplane & Whole Map & Midplane \\
        \hline
       $\log_2{(\totaldensity+1)}$ & $9.15\scinot{-2}$ & $1.35\scinot{-2}$ & $1.28\scinot{-1}$ & $4.56\scinot{-2}$ \\
       $\ionisationfraction$       & $1.71\scinot{-2}$ & $9.18\scinot{-5}$ & $3.70\scinot{-2}$ & $2.37\scinot{-3}$ \\
       $\log_2{(\etemperature+1)}$ & $7.86\scinot{-2}$ & $4.62\scinot{-6}$ & $9.23\scinot{-2}$ & $5.51\scinot{-4}$ \\
        \hline
       $\neutrals$     & $1.01\scinot{35}$ & $3.46\scinot{33}$ & $8.29\scinot{16}$ & $1.23\scinot{16}$ \\
       $\electrons$    & $7.18\scinot{35}$ & $3.95\scinot{35}$ & $3.31\scinot{17}$ & $2.20\scinot{17}$ \\
       $\etemperature$ & $6.74\scinot{0} $ & $1.11\scinot{-3}$ & $7.24\scinot{-1}$ & $7.66\scinot{-3}$ \\
       $\emission$     & $3.83\scinot{-1}$ &                   & $2.15\scinot{-1}$ &                   
    \end{tabular}
\end{table*}

\begin{table*}
    \centering
    \caption{Errors achieved by the optimisation of the parametric maps using the renders and midplane measurements, both in the parametric representation and in the SOLPS representation. MSE is the Mean Square Error and MAE is the Mean Absolute Error.}
    \label{table:recon-errors-roi-crop-diff-init}
    \renewcommand{\arraystretch}{1.5}
    \begin{tabular}{c||c|c||c|c}
        & \multicolumn{4}{c}{Errors on the Modified Experiment~\ref{experiment:all-maps-cropped-quantised} with $\ionisationfraction=0.5$ on Boundary} \\
        \cline{2-5}
        &  \multicolumn{2}{c}{MSE} & \multicolumn{2}{c}{MAE} \\
        \cline{2-5}
        & Whole Map & Midplane & Whole Map & Midplane \\
        \hline
       $\log_2{(\totaldensity+1)}$ & $1.55\scinot{-1}$ & $3.88\scinot{-2}$ & $1.62\scinot{-1}$ & $6.05\scinot{-2}$ \\
       $\ionisationfraction$       & $3.00\scinot{-2}$ & $1.18\scinot{-3}$ & $5.65\scinot{-2}$ & $8.13\scinot{-3}$ \\
       $\log_2{(\etemperature+1)}$ & $1.01\scinot{-1}$ & $9.09\scinot{-6}$ & $1.07\scinot{-1}$ & $7.63\scinot{-4}$ \\
        \hline
       $\neutrals$     & $1.66\scinot{35}$ & $2.29\scinot{34}$ & $1.19\scinot{17}$ & $3.1\scinot{16}$  \\
       $\electrons$    & $7.92\scinot{35}$ & $2.43\scinot{35}$ & $3.53\scinot{17}$ & $1.77\scinot{17}$ \\
       $\etemperature$ & $6.96\scinot{0} $ & $1.76\scinot{-3}$ & $7.72\scinot{-1}$ & $9.05\scinot{-3}$ \\ 
       $\emission$     & $3.54\scinot{-1}$ &                   & $2.04\scinot{-1}$ &                   
    \end{tabular}
\end{table*}

\end{document}